\documentclass[aps,prapplied,twocolumn,amsmath,amssymb,superscriptaddress,floatfix]{revtex4-2}

\usepackage{graphicx}	
\usepackage{amsfonts}
\usepackage{bm}         	
\usepackage{tabularx}
\usepackage{color}
\usepackage{dcolumn}		
\usepackage{calc}
\usepackage{xcolor,colortbl}
\usepackage{rotating}
\usepackage[colorlinks,citecolor=blue,linkcolor=blue]{hyperref}
\usepackage{hyperref}

\newcommand{\CNN}{Centre de Nanosciences et de Nanotechnologies, CNRS, Universit\'e Paris-Saclay, 91120 Palaiseau, France}
\newcommand{\INL}{International Iberian Nanotechnology Laboratory, Braga, Portugal}

\begin{document}

\title{
Field-driven triggering of self-induced Floquet magnons in a magnetic vortex}

\author{R. Lopes Seeger}
\email{rafael.lopes-seeger@universite-paris-saclay.fr}
\affiliation{\CNN}
\author{G. Philippe}
\affiliation{\CNN}
\author{A. Jenkins}
\affiliation{\INL}
\author{L. C. Benetti}
\affiliation{\INL}
\author{A. Schulman}
\affiliation{\INL}
\author{R. Ferreira}
\affiliation{\INL}
\author{J.-V. Kim}
\affiliation{\CNN}
\author{T. Devolder}
\affiliation{\CNN}

\date{\today}

\begin{abstract}
We report the experimental control of Floquet magnons in a magnetic vortex. Using microwave spectroscopy of vortex state magnetic tunnel junctions (MTJs), we find that self-induced Floquet sidebands form frequency combs whose existence depend on the vortex core orbit. By shifting the vortex core with an applied magnetic field, we switch the system between regular and Floquet magnons at identical drive conditions, demonstrating hysteretic control of the Floquet spectrum. 
A nonlinear vortex-magnon model shows that this behavior originates from multiple stable vortex gyration radii created by Floquet-mediated feedback. These results establish magnetic state initialization as a means to switch between regular and Floquet magnons.
\end{abstract}

\maketitle

\textit{Introduction} — Magnetic vortices in nanodisks are archetypal spin textures forming in disks of specific aspect ratios, consisting of an in-plane curling magnetization and a nanoscale out-of-plane core at their center \cite{Cowburn1999}. At remanence the vortex core is localized at the center of the disk. An in-plane magnetic field causes the vortex core to shift perpendicularly within the plane of the disk to minimize the Zeeman energy. 
Resonant excitation by a microwave field induces a gyrotropic motion of the vortex core around its equilibrium position, which typically occurs in the sub-GHz frequency range \cite{Novosad2005,Guslienko2008}. In addition to this low-frequency gyrotropic mode, there exist higher frequency spin wave modes corresponding to magnetization oscillations around the static vortex configuration \cite{Ivanov2005,Awad2010,Taurel2016}. These modes generally lie in the GHz frequency range and are geometrically quantized, reflecting the axial symmetry of the vortex ground state. They can be described by the radial and azimuthal indices $(n,m)$. Because the vortex core couples to all azimuthal spin-wave modes, vortex-state disks provide a natural platform for studying nonlinear, multimode magnon dynamics.

Under strong excitation, vortex dynamics become highly nonlinear, allowing energy to be exchanged between the vortex core and spin-wave modes \cite{Schultheiss2019,korber2020,Verba2021,jenkins_electrical_2021,wang2022,gao2023,devolder2025timeresolvedsplittingmagnonsvortex}. In this regime, interactions between the vortex core and azimuthal spin-wave modes can give rise to steady-state core gyration. This core motion induces a time-periodic modulation of the spin waves through the periodic modulation of the local effective field imposed by the vortex core’s orbit at a fixed radius, thereby renormalizing the magnon manifold and leading to the formation of Floquet sidebands \cite{heins2024selfinducedfloquetmagnonsmagnetic}. Recent theory predicts that the feedback between Floquet magnons and vortex gyration can produce multiple stable vortex orbits under the same drive conditions, leading to distinct Floquet spectra that depend on the system’s history \cite{philippe2025excitationvortexcoregyration}. Such multistability could provide a versatile control knob for tuning distinct Floquet frequency-comb spectra, complementing strategies previously reported for manipulating the Floquet modes \cite{heins2025controlmagnonfrequencycombs,heins2025coherentcontrolfloquetengineeredmagnon}. A key open question is whether such multistable Floquet states can be accessed and controlled experimentally.

While coherent control of magnon frequency combs has recently been demonstrated using short-pulse excitation schemes that drive the vortex core far from equilibrium \cite{heins2025coherentcontrolfloquetengineeredmagnon}, the present work addresses a complementary regime. Here, the vortex core is prepared in well-defined equilibrium positions. 
We demonstrate experimentally that self-induced Floquet magnons in a magnetic vortex are hysteretic and tunable through the vortex-core position. Using MTJs to electrically detect vortex dynamics, we show that displacing the vortex core with an in-plane magnetic field allows the system to switch between regular and Floquet spectra at identical microwave drive conditions. Crucially, we show that this behavior is explained by a nonlinear theory of vortex–Floquet coupling that predicts multiple stable gyration radii. Our results establish the vortex core initial position as an internal control parameter for Floquet engineering in magnetic nanostructures.

\begin{figure} 
\includegraphics[width=8.5cm]{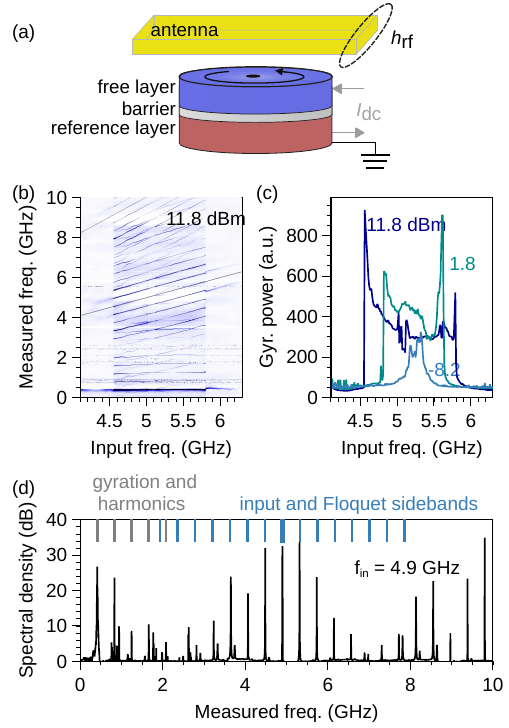}
\caption{(a) Schematic diagram showing the excitation of a MTJ using an inductive antenna. A bias current is passed through the MTJ, whose free layer hosts a magnetic vortex. The voltage fluctuations detected at the MTJ arise from the vortex core gyration, which modulates the MTJ resistance through magnetoresistive effects. The resulting signal is measured with spectrum analyzer. (b) Colormap of the power spectral density when applying an input frequency ($f_{\textrm{in}}$) in the presence of 200 mV dc bias. (c) Power of the gyration mode versus $f_{\textrm{in}}$ for different excitation powers. (d) Power spectral density presenting two sets of frequency combs, the first comprising the harmonics of the gyration frequency, and the second centered about the excitation frequency (Floquet modes).  } 
\label{Fig1}
\end{figure}

\textit{Methods} — We investigate circular magnetic elements, 45 nm thick, composed of a CoFeBSi alloy that stabilizes a vortex ground state, similar to that reported in Ref. \cite{devolder2025timeresolvedsplittingmagnonsvortex}. The MTJs exhibit a magnetoresistance ratio of about 180\% and a typical resistance of 120 $\Omega$ for 400 nm-diameter junctions, as used in the present study. The reference layer in each MTJ is uniformly magnetized and fixed in orientation, allowing vortex dynamics in the free layer to be detected electrically through resistance oscillations [Fig. \ref{Fig1}(a)] \cite{Ivanov2005,Boust2004,Devolder2017}. A 3 $\mu$m-wide inductive antenna inductive antenna positioned above the MTJ generates a microwave magnetic field that excites the vortex and spin-wave modes. The resulting voltage fluctuations across the biased MTJ are measured with a spectrum analyzer, providing direct access to both the vortex gyration frequency and the associated Floquet sidebands. 

\textit{Frequency combs and gyration radius} — We first examine the power spectrum density for input power $P_{\textrm{in}}$ = 11.8 dBm as a function of $f_{\textrm{in}}$, as shown in Fig. \ref{Fig1}(b). Several horizontal lines can be observed, the lowest frequency one corresponds to the gyrotropic frequency. Interestingly, within a certain range of drive frequencies a frequency comb emerges. Besides the frequency comb, there is a reduction of the gyration frequency ($f_{\textrm{g}} \approx $ 415 MHz), similar to reported nonlinear frequency shifts \cite{Drews2012}. 
The frequency combs lies around a spin wave mode of indices ($n = 0, m = -1$) at 4.9 GHz, which was observed by the thermal population \cite{devolder2025timeresolvedsplittingmagnonsvortex}.  
The spacing between the comb lines is set by the gyration frequency, allowing the comb frequencies to be expressed as $f = f_{\textrm{in}}+kf_{\textrm{g}}$, where $k$ is an integer. In addition, this region exhibits numerous low-frequency harmonics of $f_{\textrm{g}}$. 
The frequency combs appear only above a threshold of the input power, as observed in larger NiFe disks \cite{heins2024selfinducedfloquetmagnonsmagnetic}. By extracting signal frequency for one single $f_{\textrm{in}}$, as is shown in Fig. \ref{Fig1}(d) for $f_{\textrm{in}} = 4.9$ GHz, one can see more clearly the gyration harmonics as well as the Floquet sidebands \footnote{The gyration harmonics are flat as a function of frequency, while the Floquet sidebands are quasi-linear as a function of frequency.}. 

Figure \ref{Fig1}(c) shows the integrated power spectral density of the gyrotropic mode ($\int_{0.2~GHz}^{0.7~GHz} S(f) df $), which has been shown to scale with the vortex-core gyration radius \cite{dussaux_large_2010}, plotted as a function of the excitation frequency $f_{\textrm{in}}$ for several microwave powers. 
Outside the region where Floquet sidebands appear, the integrated power remains small, indicating that the vortex core is confined near the disk center. As $f_{\textrm{in}}$ is swept through the spin-wave resonance around 5.1 GHz, a pronounced increase in the gyration power is observed. At this frequency, the amplitude of the gyration power grows systematically with the applied microwave power, reflecting a progressive expansion of the vortex orbit. This behavior marks the transition from thermally driven Brownian motion of the core to a driven steady-state gyration sustained by nonlinear magnon–vortex coupling. In the low-power regime preceding the onset of Floquet sidebands, the MTJ detects only stochastic magnetization fluctuations, which appear as weak resistance and voltage noise and give rise to the small baseline signal in \ref{Fig1}(c). The abrupt growth of the gyration power upon entering the Floquet regime therefore provides a direct electrical signature of the emergence of a finite vortex-core orbit.

\begin{figure} 
\includegraphics[width=8.5cm]{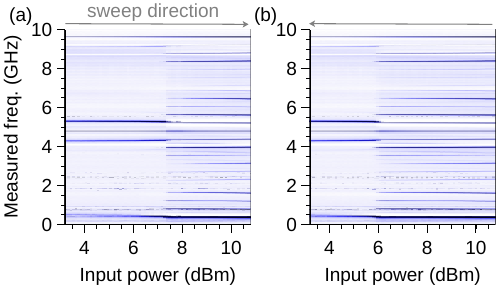}
\caption{Colormap of the power spectral density measured at a fixed excitation frequency of $f_{\textrm{in}} = 4.8$ GHz while sweeping the microwave power (a) upward and (b) downward. As the power is increased, Floquet sidebands appear only above a threshold of approximately 7.8 dBm. When the power is subsequently decreased, the Floquet spectrum persists down to about 5.8 dBm, demonstrating a clear hysteresis. In this intermediate power range, the system supports two stable steady states (one without and one with a fully developed Floquet frequency comb) revealing bistability of the driven vortex–Floquet system under identical drive conditions. 
} 
\label{FigPscan}
\end{figure}

\textit{Hysteresis in power sweeps} — Figure \ref{FigPscan} shows the power spectral density measured at a fixed excitation frequency while the microwave power is swept from low to high [Fig. \ref{FigPscan}(a)] and from high to low [Fig. \ref{FigPscan}(b)]. As the input power is increased, the directly driven mode and the associated microwave mixing products at $f_{\textrm{in}}\pm f_{\textrm{g}}$ appear at 7.8 dBm, consistent with previous observations of nonlinear vortex–magnon coupling \cite{devolder2025timeresolvedsplittingmagnonsvortex}. Upon further increasing the power, a full Floquet frequency comb develops, characterized by multiple sidebands spaced by the gyration frequency.

When the power is subsequently reduced, the Floquet spectrum persists down to approximately 5.8 dBm, well below the threshold at which it first appeared. This demonstrates a clear hysteresis in the formation of the Floquet states. In this intermediate power range, two distinct steady states coexist: one in which the vortex core remains localized near the disk center and no Floquet sidebands are present, and another in which the vortex executes a large-amplitude orbit that sustains a full Floquet frequency comb. It is noteworthy that the hysteresis emerges only within a finite window of excitation frequencies  $f_{\textrm{in}}$, outside of which the system supports a single stable gyration state. A non-hysteretical case for a different $f_{\textrm{in}}$ is shown in the supplementary material \cite{SM} 

As discussed below, this bistability originates from the nonlinear feedback between the vortex-core gyration and the Floquet modes, which creates distinct stable gyration radii under identical drive conditions. Because the realized steady state depends on the system’s history, Floquet magnons can be generated below the spontaneous instability threshold if the vortex core is first displaced from its equilibrium position. In the following section, we exploit this property by actively shifting the vortex core with an applied magnetic field.

\begin{figure} 
\includegraphics[width=8.5cm]{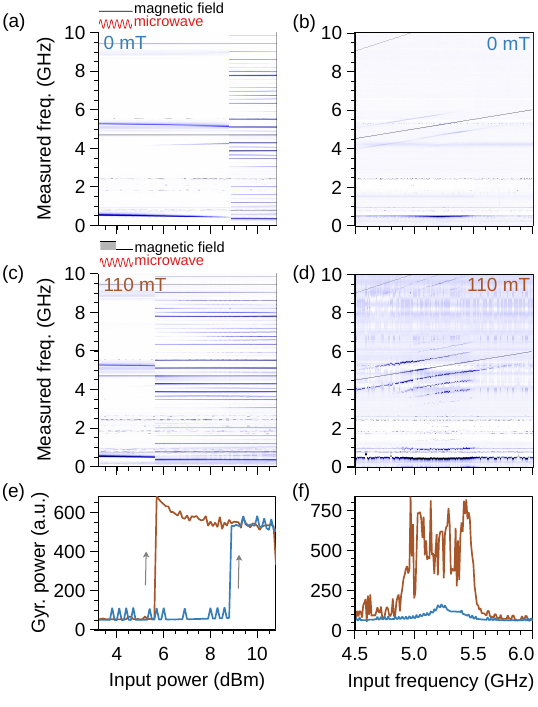} 
\caption{Colormaps of the power spectral density as a function of microwave power measured at a fixed excitation frequency of $4.7$ GHz for two different vortex-core initializations: vortex centered (a) and vortex displaced by an in-plane magnetic field (c). The magnetic-field-initialized state exhibits Floquet sidebands at significantly lower drive powers, demonstrating a history-dependent instability threshold. Power spectral density as a function of excitation frequency at a fixed input power of $-3.2$ dBm for the same two vortex-core initializations: vortex centered (b) and vortex displaced (d). Floquet sidebands are absent when the core is initially centered but appear when the core is displaced by the magnetic field, although all other drive conditions are identical. Each spectrum is measured after an independent initialization with the in-plane magnetic field reset to zero. (e) Integrated power of the gyrotropic mode as a function of microwave power for the two initial states, showing that magnetic-field displacement increases the steady-state gyration radius and lowers the instability threshold. (f) Corresponding integrated gyration power versus excitation frequency, confirming that a larger gyration radius promotes the emergence of Floquet sidebands.} 
\label{FigHpulse}
\end{figure}
\textit{Stimulation with a magnetic field} — To probe the role of the initial vortex-core position and uncover hysteretic effects in the driven vortex-Floquet system, we employ an in-plane magnetic field to controllably displace the vortex core from the disk center prior to the application of the microwave drive. This procedure prepares the system in well-defined initial equilibrium states that differ in their core position, allowing access to distinct stable gyration radii under otherwise identical excitation conditions. After initialization, the microwave field is switched on, the in-plane magnetic field is brought back to zero, and the vortex relaxes into a steady-state gyrotropic orbit whose radius depends on the prior displacement history. Figure \ref{FigHpulse} shows the resulting power spectral density measured as a function of microwave power at fixed frequency (a,c) and as a function of microwave frequency at fixed power (b,d) for different initialization fields. By comparing spectra obtained from a centered ground state and from displaced equilibrium configurations, we demonstrate that a single driving frequency and power can give rise to distinct spectral responses, reflecting the bistability of the nonlinear vortex dynamics. %

The top panel of Fig. \ref{FigHpulse}(a) shows that when the vortex core is initially centered, Floquet sidebands appear only above a threshold of approximately 9 dBm. In contrast, when the core is displaced by the applied magnetic field (c), Floquet sidebands emerge at a significantly lower threshold of about 5.5 dBm. In addition, the gyration frequency $f_{\textrm{g}}$ changes by roughly $\pm 50$ MHz upon application of the field, reflecting pinning and material granularity effects \cite{Compton2006, jenkins_impact_2024}, and not a nonlinear frequency shift, since the change occurs already at low excitation amplitudes and correlates with the static core displacement. This shift in $f_{\textrm{g}}$ is directly imprinted on the Floquet spectra and is visible in all measurements performed after magnetic-field initialization.

A similar history dependence is observed when sweeping the microwave frequency, as shown in Fig. \ref{FigHpulse}(b,d). When the vortex core starts from the disk center, no Floquet sidebands are observed in this power range. However, after displacing the core with a magnetic field, pronounced Floquet sidebands develop over the same frequency interval, demonstrating that the Floquet spectrum depends sensitively on the vortex-core initial position.

To clarify the origin of this behavior, we extract the integrated power of the gyrotropic mode, which is proportional to the vortex-core gyration radius. Figure \ref{FigHpulse}(c) shows that the applied magnetic field increases the steady-state gyration radius and lowers the threshold for sustained motion. When the microwave frequency is swept, this enhanced gyration radius promotes the formation of Floquet sidebands at lower excitation levels, as shown in Figure \ref{FigHpulse}(d). These results demonstrate that the magnetic field acts as an internal control parameter that selects distinct Floquet states by setting the initial vortex-core position.



\textit{Discussion} — Our experimental results are interpreted in terms of the theoretical model developed in Ref. \cite{philippe2025excitationvortexcoregyration}. This model describes the relation between the gyration radius and Floquet modes. Nonlinear interactions of Floquet modes with the gyration mode modifies the gyration radius and the susceptibility of Floquet modes is a function of gyration radius. To quantify the effect of these nonlinear interaction on the gyration radius, we run simulations to extract susceptibilities of Floquet modes depending of the gyration radius. In these simulations, a rotating field at the gyration frequency maintains the gyration radius constant, then a low power in-plane field at $f_{\textrm{rf}}$ activates the Floquet modes without affecting the gyration radius. The amplitude of the rotating field is gradually increased to obtain the value of susceptibilities in a chosen gyration radius range. The susceptibility of a Floquet mode is evaluated using the power of this mode divided by the in-plane field power. The last step is to sum susceptibility of all modes that contribute to increase the gyration radius minus the sum of susceptibilities of all modes that contribute to reduce the gyration radius, we call this quantity the nonlinear interaction contribution, given by: \\
$$f(R) \propto  B_{rf} \omega_{g} \sum_{k=1}^{N} C_{-k} a_{-k}a_{-k+1}-C_{k} a_{k}a_{k-1},  $$
where $\omega_g$ is the gyration frequency, $C_k$ are the coupling constant, and $a_k$ are the modes susceptibility.
By accounting the nonlinear interaction contribution in the Thiele model \cite{Guslienko2008}, we obtain:
$$ \dot{R}=-\Gamma_{g} R + f(R),$$ 
where $\Gamma_{g} R$ is the relaxation function. 
Consequently, the specific form of the function $f(R)$ can allow the system to support nontrivial steady-gyration states. These correspond to equilibria with $\dot{R}=0$ that satisfy $\Gamma_g R_0 = f(R_0)$, where the equilibrium radius $R_0 \neq 0$.

\begin{figure} 
\includegraphics[width=8.5cm]{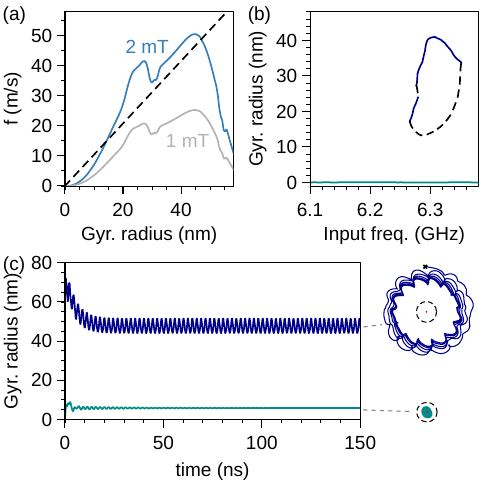} 
\caption{(a) Nonlinear interaction contribution to the gyration, $f(R_{g})$, as a function of the steady-state gyration radius for a driving frequencies $f_\textrm{in}$ = 6.3 GHz. $f(R_\textrm{g})$ is obtained for two different rf drive field amplitudes. The dashed line represents the relaxation function $\Gamma_\textrm{g} R_\textrm{g}$. (b) $R_\textrm{g}$ as a
function of driving field frequency when $B_{\textrm{rf}}$ = 2 mT. Solid lines represent stable fixed points and dashed lines represent unstable fixed points. (c) Time evolution of the gyration radius for two simulations initialized at different vortex core positions. } 
\label{FigModelling}
\end{figure}

On Fig. \ref{FigModelling}(a), we see the nonlinear interaction contribution for an in plane rf field with a frequency of 6.3 GHz and an amplitude of 1 mT and 2 mT. It is linearly dependent on the amplitude of the in plane field. 
When we compare the nonlinear interaction contribution to the relaxation function, represented by the dashed line on \ref{FigModelling}(a), we see than at 1 mT both terms are equal only when the gyration radius is equal to 0 nm. However, at 2 mT there are two other positions where both terms are equal. These are the equilibrium gyration radii. If the radius remains unchanged after a small perturbation, the gyration radius is stable. If a small perturbation causes the radius to change significantly, the gyration radius is unstable. We notice that on Fig. \ref{FigModelling}(a) there is only one stable gyration radius at 0 nm for an amplitude of 1 mT, but there is two stable radius at 0 nm and 40 nm for an amplitude of 2 mT.
The nonlinear interaction contribution $f(R)$ also depends on the excitation frequency $f_{\textrm{rf}}$. In Fig. \ref{FigModelling}(b) we can see stable equilibrium radii, represented by solid lines, and unstable equilibrium radii, represented by dashed lines, depending on $f_{\textrm{rf}}$ when $B_{\textrm{rf}}$ = 2 mT. The bistable radii exist only on a specific frequency range narrower than 100 MHz, for other frequencies the only stable equilibrium position is 0 nm. The evolution of gyration radius in a configuration with bistable radius depends of its initial position. 
The time evolution of the gyration radius, resulting from two different simulations with distinct initial conditions for the gyration radius are shown in Fig. \ref{FigModelling}(c), where the vortex core positions are defined by an applied dc in-plane field (20 mT). They further demonstrate that the final steady state depends on the initial vortex-core position: if the core starts beyond the unstable fixed point, it evolves toward the finite-radius orbit that sustains a Floquet frequency comb, otherwise it will relax to the disk center and no Floquet state forms.

This model provides a direct interpretation of the experimentally observed hysteresis and history dependence. At low microwave power, the only stable solution is $R=0$, corresponding to a vortex core localized at the disk center and the absence of Floquet sidebands. At high power, a finite-radius orbit becomes the only stable state, leading to the formation of a Floquet frequency comb. In an intermediate range of drive powers, both solutions coexist, giving rise to bistability. When the power is increased, the system remains in the $R=0$ state until it loses stability and jumps to the finite-radius orbit. When the power is decreased, the system stays on the finite-radius branch until that state becomes unstable. This explains the hysteresis observed [Fig. \ref{FigPscan}] and why magnetic-field displacement of the vortex core [Fig. \ref{FigHpulse}] enables access to Floquet states below the spontaneous instability threshold.

\textit{Conclusion} — We have investigated the complex nonlinear interactions between magnon modes and vortex core gyration, which give rise to magnon Floquet states, using microwave electrical measurement in MTJs. In particular, our study reveals how the dynamic motion of the vortex core plays a crucial role in determining the threshold conditions for the emergence of a Floquet spectrum. These results highlight the strong coupling between spin-wave dynamics and vortex motion, emphasizing the importance of nonlinear effects in such systems. The sensitivity of the Floquet spectra to magnetic-field history further highlights the role of nonlinear effects. 

\textit{Acknowledgments} — We acknowledge financial support from the EU Research and Innovation Programme Horizon Europe under grant agreement n°101070290 (NIMFEIA).

\bibliography{scibib}

\clearpage

\end{document}